\documentclass{article}
\usepackage{natbib}
\usepackage[final]{neurips_2024}
\usepackage[utf8]{inputenc} %
\usepackage[T1]{fontenc}    %
\usepackage[hidelinks]{hyperref}       %
\usepackage{url}            %
\usepackage{booktabs}       %
\usepackage{amsfonts}       %
\usepackage{nicefrac}       %
\usepackage{microtype}      %
\usepackage{xcolor}         %
\usepackage{graphicx}
\usepackage{longtable}
\usepackage{caption}
\usepackage{mdframed}
\usepackage{subcaption}
\usepackage{multirow}
\usepackage{placeins}
\usepackage{multicol}
\usepackage{makecell}
\usepackage[normalem]{ulem} %
\usepackage{wrapfig}
\usepackage[percent]{overpic}
\usepackage{lipsum}
\usepackage{csquotes}
\usepackage[OT2,T1]{fontenc}
\usepackage[english]{babel}
\usepackage{devanagari}
\usepackage{tablefootnote}
\usepackage{pdfpages}
\usepackage{caption}
\usepackage{amsmath}
\usepackage{enumitem}
\usepackage{changepage}

\usepackage{macros}

\usepackage{hyperref}  
\hypersetup{
     colorlinks=true,
     linkcolor=blue,
     citecolor=blue,
     filecolor=blue,
     urlcolor=blue,
 }

\title{Seed-Music: A Unified Framework for High Quality and Controlled Music Generation}

\author{Seed Team, ByteDance\thanks{Please cite this work as ``Seed-Music (2024)''. The full statement of author contributions and acknowledgments can be found at the end of the document. Correspondence regarding this technical report should be sent to \href{mailto:Seed-Music@bytedance.com}{Seed-Music@bytedance.com}.}}

\begin{document}
% \commenttrue

\maketitle
\begin{abstract}
We introduce Seed-Music, a suite of music generation and editing systems designed to produce high-quality music with fine-grained style control. Our unified framework leverages both auto-regressive language modeling and diffusion approaches to support two key music creation workflows: \textit{controlled music generation} and \textit{post-production editing}. For controlled music generation, our system enables vocal music generation with performance controls from multi-modal inputs, including lyrics, style descriptions, audio references, musical scores, and voice prompts. For post-production editing, it offers interactive tools for editing vocal lyrics, melodies, and timbres directly within an existing music audio track.
We encourage readers to explore the demo audio examples at \url{https://team.doubao.com/seed-music}.
\end{abstract}

%%% INTRODUCTION %%%
\section{Introduction}
\label{sec:intro}

Music is deeply embedded in human culture. Throughout human history, vocal music has accompanied key moments in life and society: from love calls to seasonal harvests \citep{UNIVERSALITY-HUMAN-MUSIC}. Today, vocal music remains central to global culture. However, creating vocal music is a complex, multi-stage process involving pre-production, writing, recording, editing, mixing, and mastering \citep{owsinski2010music,senior2012mixing}, making it challenging for most people. 
Our goal is to leverage modern generative modeling technologies, not to replace human creativity, but to lower the barriers to music creation. By offering interactive creation and editing tools, we aim to empower both novices and professionals to engage at different stages of the music production process.

Today, deep generative models are capable of understanding and generating multi-modal data \citep{MULTIMODAL-LLM-SURVEY-yin2023survey, ANYGPT-zhan2024anygptunifiedmultimodalllm}. Although music generation has benefited from the advances in natural language processing \citep{GPT2-radford2019language}, speech synthesis \citep{SEED-TTS-anastassiou2024seedttsfamilyhighqualityversatile}, and computer vision \citep{DIFFUSION-peebles2023scalablediffusionmodelstransformers}, the task remains challenging for several reasons:
\begin{itemize}[leftmargin=2em]

%\item \textbf{Domain Complexity} Music signals are complex. Compared to speech and environmental sounds, instrumental music signals can contain multiple overlapping tracks, rich tones and timbres as well as a wide bandwidth of frequencies. From a semantic standpoint, most music signals contain short term melodic coherency and long term structural coherency. Compared to text-to-speech (TTS) tasks, singing vocals are much wider in pitch range and feature many more singing styles and expressive techniques. Vocal music generation arguably combines the most difficult aspects of instrumental, environmental and speech sounds. The generation output must simultaneously produce intelligible words and melodic singing, produce harmonic tones for melodies and chords, produce rhythmic and transient sounds like percussion and effects whilst being coherent across the entire 3 minutes of a full length track. Despite recent breakthroughs in generative modeling, simply scaling up the model size and swapping speech data with music data cannot guarantee promising musical compositions and sounds. There is still much room for improvement in terms of musicality and audio quality.

\item \textbf{Domain complexity}: Music signals are highly complex, exhibiting both short-term melodic coherence and long-term structural consistency. Unlike speech, vocal music features overlapping sounds across a wide frequency range. Singing, with its wide pitch range and expressive techniques, adds another layer of complexity. The model must simultaneously generate melodic vocals, harmonic tones, and rhythmic percussion.

\item \textbf{Evaluation difficulty}: Evaluating music generation models often requires domain expertise to assess artistic quality. This includes judging the appeal of melodies, the coherence of chord progressions, the presence of idiomatic structure, and the expressiveness of vocals. Many of these aesthetic qualities are deeply influenced by cultural and regional differences. Quantifying these artistic elements in music poses a challenge.

%\item \textbf{Data Limitations} The quantity of recorded music data is significantly less than speech. Moreover, there is a much higher barrier to producing a music dataset covering a wide variety of instrumentation and genres compared to producing a speech dataset of different languages. 
%
%In addition to the scarcity of music audio data, annotating music audio is also a huge challenge. In order for the generative model to learn to steer generation according to conditions such as lyrics, genre, instruments, and song structure, we need paired training data of music audio and its annotations. One way to obtain such annotations is through using content-based Music Information Retrieval (MIR) models \citep{casey2008content} to label the music audio automatically, yet training MIR models also requires large quantities of annotated data.
%
%Alternatively, we can directly obtaining music annotation with human labeling, which requires specialized domain knowledge. Most individuals can transcribe speech or label an image, but labelling a bar, chord, or section tag requires musical background. Many labelled musical datasets such as \cite{goto2002rwc}, \cite{bertin2011million}, \cite{nieto2019harmonix}, \citet{LAKH-DATASET-raffel2016learning, MAESTRO-DATASET-hawthorne2018enabling}, and \cite{wang2020pop909} are orders of magnitudes smaller than their image and speech counterparts. 

\item \textbf{Data complexity}: 
Generative models require annotated music data to learn how to condition outputs based on lyrics, genre, instrumentation, and song structure. However, music annotations require specialized domain knowledge. While tasks like speech transcription or image labeling are accessible to many, identifying musical elements such as chords, song sections, instruments, and genres requires a strong musical background. 
%Moreover, automatically labeling music audio with content-based Music Information Retrieval (MIR) models \citep{casey2008content} also depends on large amounts of annotated data to effectively train these systems.
%For generative models to learn how to condition outputs on lyrics, genre, instrumentation, and song structure, annotated music data is essential. Music annotation requires special domain knowledge. While many people can transcribe speech or label images, identifying musical elements such as chords, song sections, instruments, and genres requires a strong musical background. Automatically labeling music audio with content-based Music Information Retrieval (MIR) models \citep{casey2008content} also requires large quantities of annotated data to train the models.

\item \textbf{Diverse user segments and needs}: 
The needs of novice creators differ greatly from those of professionals. A text-to-music system \citep{MUSICLM-agostinelli2023musiclmgeneratingmusictext, MUSICGEN-copet2024simplecontrollablemusicgeneration} that generates a full audio piece from a text prompt can be transformative for a beginner, but may offer limited values to professional producers, who often seek more granular control over compositions and access to individual instrument stems. Even among professionals, needs differ: a guitarist might need vocal editing tools, while a vocalist might want to tweak guitar or other instrument tracks.
\end{itemize}

\paragraph{Contributions}
With these challenges in mind, we highlight the versatility of Seed-Music. It supports vocal and instrumental music generation, singing voice synthesis, singing voice conversion, music editing and more. Our methodology, experiments, and solutions are designed to address diverse use cases. Rather than relying on a single modeling approach, such as auto-regression (AR) or diffusion, we propose a unified framework that adapts to the evolving workflows of musicians.

Our key contributions are threefold:
\begin{itemize}[leftmargin=2em]
    \item We introduce a unified framework that leverages both auto-regressive language modeling and diffusion approaches for high-quality vocal music generation conditioned on diverse and multi-modal inputs.
    \item We present a diffusion-based approach that enables fine-grained editing of music audio.
    \item We propose a novel zero-shot singing voice conversion method, which requires only a 10-second recording of either singing or speech from the user.
\end{itemize}

In Section~\ref{sec:literature}, we give a brief literature review of music generation. In Section~\ref{sec:method}, we introduce our unified framework,
which is built upon three fundamental representations: audio tokens, symbolic tokens, and vocoder latents. The corresponding pipelines and design choices will be detailed. In Section~\ref{sec:experiments}, we dive into how our unified framework can be configured and trained to support various music generation and editing tasks. In Section~\ref{sec:conclusion} and Section~\ref{sec:ethics-safety}, we discuss potential applications and limitations of Seed-Music, including those related to building safe and ethical generative AI systems.

%consisting of three parts: (1) a representation learning module that converts raw audio into a highly compressed representation, (2) a generation module that predicts the representation based on various user controls, and (3) a rendering module trained to produce the waveform from the representation. 

%%% LITERATURE REVIEW %%%%
\section{Literature Review}
\label{sec:literature}

Vocal music is typically defined as the combination of a vocal track and an instrumental track. Historically, these tracks were generated separately and then mixed together to create the full mixture. With advances in machine learning based generative modeling, Jukebox \citep{JUKEBOX-dhariwal2020jukeboxgenerativemodelmusic} was one of the first systems capable of generating full vocal music mixes in an end-to-end fashion. While its generation was prohibitively slow, Jukebox has demonstrated the ability to produce vocal music that aligned closely with input lyrics, as well as specified artists and genres. Today, many AI music creation platforms enable creators to instantly generate vocal music on demand using natural language prompts.
The field of music generation has a long history and can be broadly categorized into two areas: symbolic domain and audio domain generations. This section covers a number of important works representing both approaches.
%The field of music generation has a long history. Music generation systems can largely be divided into two categories: symbolic domain and audio domain generations. A number of representative works are covered in this section. 

\paragraph{Symbolic music based systems.} 
%The earliest instrumental music generation systems were rule-based and generated notes in limited styles \citep{EMI}. These systems generate symbolic forms, which is then rendered into audio using digital signal processing (DSP) methods. Later, rule-based approaches were replaced with data-driven approaches for music generation. Examples include FolkRNN \citep{FOLKRNN-sturm2016musictranscriptionmodellingcomposition}, PerformanceRNN \citep{PERFORMANCE-RNN-2017}, MusicTransformer \citep{MUSIC-TRANSFORMER-huang2018musictransformer}, MuseNet \citep{MuseNet}, CoCoNet \citep{COCONET-huang2019counterpointconvolution} and many others. Although many systems relied mainly on MIDI (Music Instrument Digital Inferfaces) data, other symbolic notations such as ABC notation, tablature and music XML \citep{MUSPY-dong2020muspy, DADAGP-GUITARTABS-sarmento2021dadagp} were also explored. The defining limitation for all these systems was the limited availability of training data. MIDI transcriptions are notoriously difficult, time-consuming and expensive to procure. 
Early instrumental music generation systems were rule-based, generating notes in limited styles \citep{EMI}. These systems generated symbolic forms, which were then rendered into audio using digital signal processing (DSP) methods. Over time, rule-based approaches were replaced with data-driven ones for music generation. Examples include FolkRNN \citep{FOLKRNN-sturm2016musictranscriptionmodellingcomposition}, PerformanceRNN \citep{PERFORMANCE-RNN-2017}, MusicTransformer \citep{MUSIC-TRANSFORMER-huang2018musictransformer}, MuseNet \citep{MuseNet}, CoCoNet \citep{COCONET-huang2019counterpointconvolution}, and others. While many of these systems primarily relied on MIDI (Music Instrument Digital Interface) data, other symbolic notations such as ABC notation, tablature, and MusicXML were also explored \citep{MUSPY-dong2020muspy, DADAGP-GUITARTABS-sarmento2021dadagp}. A major limitation of these systems was the scarcity of training data, as MIDI transcriptions are notoriously difficult, time-consuming, and expensive to produce.

\paragraph{Audio rendering for symbolic music based systems.}
%A symbolic system, whether rule-based or data-driven, requires an audio renderer to produce sounds. Traditional audio renderers employ DSP-based or sample-bases synthesizers to render instrumental sounds. A parallel line of research applied a data-driven approach to audio synthesis, which is often referred to as ``neural audio synthesis''. Examples include NSynth \citep{NSYNTH-engel2017neuralaudiosynthesismusical}, GANSynth \citep{GANSYNTH-engel2019gansynthadversarialneuralaudio}, DDSP \citep{DDSP-engel2020ddspdifferentiabledigitalsignal}, MIDI-DDSP \citep{MIDI-DDSP-wu2022mididdspdetailedcontrolmusical}, WAVERNN-based approaches \citep{WAVERNN-kalchbrenner2018efficientneuralaudiosynthesis, FAST-FLEXIBLE-SYNTHESIS}. These techniques opened up new timbral and tonal possibilities compared to traditional approaches \citep{SERRA-SMITH}. For example, a WaveNet-based model can render expressive and nuanced piano recordings from generated piano scores \citep{WAV2MIDI2WAV-hawthorne2018enabling}. In the vocal generation domain, a Singing Voice Synthesis (SVS) can render vocal performances based on notes and lyric phonemes generated by a symbolic system \citep{SVS-HISTORY-cook1996singing, EARLY-SVS-DEEP-LEARNING-nishimura2016singing, yi2019singing, SVS-XIAOICESING-lu2020xiaoicesing, zhuang2021litesing}. 
% Early neural audio synthesis models did not support vocal generation since they were trained on instrumental music (e.g. Solo piano instrumentals only as in \cite{MAESTRO-DATASET-hawthorne2018enabling}).
Whether rule-based or data-driven, symbolic music generation systems require an audio renderer to produce sounds. Traditional audio renderers use DSP-based or sample-based synthesizers to render instrumental sounds. A parallel line of research has applied data-driven approaches to audio synthesis, often referred to as ``neural audio synthesis''. Notable examples include NSynth \citep{NSYNTH-engel2017neuralaudiosynthesismusical}, GANSynth \citep{GANSYNTH-engel2019gansynthadversarialneuralaudio}, DDSP \citep{DDSP-engel2020ddspdifferentiabledigitalsignal}, MIDI-DDSP \citep{MIDI-DDSP-wu2022mididdspdetailedcontrolmusical}, and WaveRNN-based approaches \citep{WAVERNN-kalchbrenner2018efficientneuralaudiosynthesis, FAST-FLEXIBLE-SYNTHESIS}. These techniques have expanded the range of timbral and tonal possibilities beyond traditional approaches \citep{SERRA-SMITH}. For instance, WaveNet-based models can render expressive and nuanced piano performances from generated scores \citep{WAV2MIDI2WAV-hawthorne2018enabling}.
In the vocal generation domain, Singing Voice Synthesis (SVS) systems render vocal performances based on notes and lyric phonemes generated by symbolic music systems \citep{SVS-HISTORY-cook1996singing, EARLY-SVS-DEEP-LEARNING-nishimura2016singing, yi2019singing, SVS-XIAOICESING-lu2020xiaoicesing, zhuang2021litesing}.

\paragraph{Language model based generative approaches.} 
Recent advancements in language modeling have introduced a new paradigm for generating musical audio. AudioLM \citep{AUDIOLM-borsos2023audiolmlanguagemodelingapproach} demonstrated that waveform generation can be framed as a next-token prediction task, where tokens are extracted from a neural codec such as SoundStream \citep{zeghidour2021soundstream}. In this work, our approach is inspired by LM-based systems in speech synthesis, including Seed-family models \citep{SEED-TTS-anastassiou2024seedttsfamilyhighqualityversatile, SEED-ASR-bai2024seedasrunderstandingdiversespeech}, VALLE-style models \citep{VALLE-wang2023neuralcodeclanguagemodels, VALLEX-zhang2023speakforeignlanguagesvoice, VALLE2-chen2024valle2neuralcodec}, and others \citep{betker2023better, lajszczak2024base}. 

%In music generation, LM-based modeling marks a paradigm shift. MusicLM \citep{MUSICLM-agostinelli2023musiclmgeneratingmusictext} could train directly on full mixes of music at unprecedented scales. These new end-to-end systems are composed of a hierarchy of modules that blur the historical separation of symbolic generation and audio synthesis. From a high level, the singing content and the music style are captured by ``semantic tokens'', which are manipulated by LMs at earlier stages of the hierarchy, while instrumental timbre and high frequency details are captured by ``acoustic tokens'' and manipulated in later stages of the hierarchy.
In music generation, LM-based approaches have shown great promise. MusicLM \citep{MUSICLM-agostinelli2023musiclmgeneratingmusictext} introduced a framework for training directly on full music mixes at an unprecedented scale. This new end-to-end system integrates modules in a hierarchical structure that blurs the traditional distinction between symbolic music generation and audio synthesis. At a higher level, ``semantic tokens'' capture elements such as vocal content and musical styles, which are modeled by LMs in the earlier stages of the hierarchy. In the later stages, ``acoustic tokens'' are handled to characterize timbral attributes and high-frequency nuances of audio. 

%Additionally, an LM-based approach enables other modalities like text to be effectively transformed into music. Instead of procuring an impossible amount of paired text-music data, MusicLM circumvented this need by extracting semantic tokens from pre-trained models like MuLan \citep{huang2022mulan} and Wav2Vec-BERT \citep{WAV2VEC-BERT-chung2021w2vbertcombiningcontrastivelearning}. An LM-based model then transforms these semantic tokens into target audio tokens representing music. Follow-up works usually have a similar high level recipe: a neural audio codec tokenizes the audio into discrete codes, a GPT2-like system \citep{vaswani2017attention, GPT2-radford2019language, touvron2023llama} performs next-token prediction of audio tokens based on multi-modal inputs that are also casted into token format, and the neural audio codec futher renders the waveform based on the predicted token sequence. These methods can also be applied to singing vocals \citep{SVC-LM-li2024self, SVS-LM-li2024accompaniedsingingvoicesynthesis, MVoice-SVS-LM--huang2024mvoice}.
% a pretrained language model in another modality captures essential semantics from that domain (usually text \citep{NLP-T5-raffel2023exploringlimitstransferlearning, NLP-FLAN-T5-chung2022scalinginstructionfinetunedlanguagemodels, CLAP-elizalde2022claplearningaudioconcepts}, but could also be images, video or other control signals) and

LM-based approaches also enable the incorporation of different modalities, such as text, into music generation. MusicLM utilizes semantic tokens extracted by pre-trained models like MuLan \citep{huang2022mulan} and Wav2Vec-BERT \citep{WAV2VEC-BERT-chung2021w2vbertcombiningcontrastivelearning}, reducing the need for vast amounts of paired text-music data. Based on these semantic tokens, an auto-regressive LM generates audio tokens that represent music audio. Follow-up approaches typically adhere to this pipeline: a neural audio codec tokenizes audio signals into discrete codes, a GPT-2-style LM \citep{vaswani2017attention, GPT2-radford2019language, touvron2023llama} predicts a sequence of audio tokens from multi-modal inputs, and a neural audio codec renders the final waveform from the predicted tokens. These methods have also been successfully applied to singing vocals \citep{SVC-LM-li2024self, SVS-LM-li2024accompaniedsingingvoicesynthesis, MVoice-SVS-LM--huang2024mvoice}.

\paragraph{Diffusion-based generative models.} 
%An explosion of diffusion-based systems are revolutionizing the way images, video and audio are generated. We draw inspirations from the vision field (\citep{DDPM-ho2020denoisingdiffusionprobabilisticmodels, DDIM-song2022denoisingdiffusionimplicitmodels, SCORE-song2021scorebasedgenerativemodelingstochastic, LATENT-DIFFUSION-rombach2022highresolutionimagesynthesislatent}) and adapt these for music generation.
The emergence of diffusion-based models has revolutionized the generation of images, video, and audio. Inspired by recent advances in the vision field \citep{DDPM-ho2020denoisingdiffusionprobabilisticmodels, DDIM-song2022denoisingdiffusionimplicitmodels, SCORE-song2021scorebasedgenerativemodelingstochastic, LATENT-DIFFUSION-rombach2022highresolutionimagesynthesislatent}, we adapt these techniques for music generation.

%For instrumental music generation, a variety of works successfully implemented text-to-music based on a diffusion backbone: Noise2Music \citep{NOISE2MUSIC-huang2023noise2musictextconditionedmusicgeneration}, Stable Audio \citep{STABLEAUDIO-evans2024fasttimingconditionedlatentaudio}, Stable Audio Open \citep{STABLEAUDIO-OPEN-evans2024stableaudioopen}, MUSTANGO \citep{MUSTANGO-DIFFUSION-melechovsky2024mustangocontrollabletexttomusicgeneration} and MusicLDM \citep{MUSICLDM-chen2023musicldmenhancingnoveltytexttomusic, AUDIOLDM2-liu2024audioldm2learningholistic}. 
%From a high level, instead of modeling high-dimensional waveforms directly in the diffusion process, these recipes often employ latent diffusion \citep{LATENT-DIFFUSION-rombach2022highresolutionimagesynthesislatent} on normalized latents extracted from a vocoder that was trained separately. The process of denoising the intermediate representation can then be conditioned on various music annotations to support fine-grain and controlled generation. These include contour-based and downbeat conditioning in Music-ControlNet \citep{MUSIC-CONTROLNET-wu2023musiccontrolnetmultipletimevarying}, chroma-based melody in Music-Gen \citep{MUSICGEN-copet2024simplecontrollablemusicgeneration} or mixtures of the above in JASCO \citep{JASCO-tal2024jointaudiosymbolicconditioning}. 
For instrumental music generation, several works have successfully implemented text-to-music generation based on diffusion backbones. Examples include Noise2Music \citep{NOISE2MUSIC-huang2023noise2musictextconditionedmusicgeneration}, Stable Audio \citep{STABLEAUDIO-evans2024fasttimingconditionedlatentaudio}, Stable Audio Open \citep{STABLEAUDIO-OPEN-evans2024stableaudioopen}, MUSTANGO \citep{MUSTANGO-DIFFUSION-melechovsky2024mustangocontrollabletexttomusicgeneration}, and MusicLDM \citep{MUSICLDM-chen2023musicldmenhancingnoveltytexttomusic, AUDIOLDM2-liu2024audioldm2learningholistic}.
Rather than directly modeling raw waveforms in the diffusion process, these methods often use latent diffusion \citep{LATENT-DIFFUSION-rombach2022highresolutionimagesynthesislatent} on normalized latents extracted from a separately trained vocoder. The denoising process on these intermediate representations (or latents) can be conditioned on various music attributes to enable fine-grained and controlled generation. Examples include the time-varying controls conditioning (e.g. dynamic, melody, rhythm) in Music-ControlNet \citep{MUSIC-CONTROLNET-wu2023musiccontrolnetmultipletimevarying}, chroma-based melody conditioning in Music-Gen \citep{MUSICGEN-copet2024simplecontrollablemusicgeneration}, and combinations of these in JASCO \citep{JASCO-tal2024jointaudiosymbolicconditioning}.

%In the vision domain, diffusion-based systems have generalized to tasks beyond generation from scratch. These include in-painting, style transfer and contour-based editing \citep{SDEDIT-meng2021sdedit, DIFFEDIT-CV-couairon2022diffeditdiffusionbasedsemanticimage, DIFFUSION-IMG2IMG-su2023dualdiffusionimplicitbridges, DIFFUSION-IMAGE-TEXT-kawar2023imagictextbasedrealimage, STABLE-DIFFUSION-3-esser2024scalingrectifiedflowtransformers}. We note many exciting analogues in the vocal music domain. For example, recent singing voice synthesis (SVS) \citep{DIFFSINGER-liu2022diffsingersingingvoicesynthesis} and singing voice conversion (SVC) \citep{SVC1-li2024real, SVC2-chen2024ldm, SVC3-yamamoto2023comparative} approaches can be thought of as a type of controllable style transfer.
% while disentangled vocal and instrumental editing can be thought of as inpainting.
In the vision domain, diffusion-based applications have expanded beyond pure generation to tasks like in-painting, style transfer, and contour-based editing \citep{SDEDIT-meng2021sdedit, DIFFEDIT-CV-couairon2022diffeditdiffusionbasedsemanticimage, DIFFUSION-IMG2IMG-su2023dualdiffusionimplicitbridges, DIFFUSION-IMAGE-TEXT-kawar2023imagictextbasedrealimage, STABLE-DIFFUSION-3-esser2024scalingrectifiedflowtransformers}. We see exciting similarities in the vocal music domain. Recent approaches for SVS \citep{DIFFSINGER-liu2022diffsingersingingvoicesynthesis} and singing voice conversion \citep{SVC1-li2024real, SVC2-chen2024ldm, SVC3-yamamoto2023comparative} can be considered as a form of controllable style transfer, much like in visual applications.

\textbf{Representation learning.} 
%In both LM-based and diffusion-based generative modeling, a critical component is the choice of representation for the audio signal. Methods for encoding speech, environmental sound and music exist on a spectrum ranging from semantic representations to acoustic representations. On one end, CLAP \citep{CLAP-elizalde2022claplearningaudioconcepts} and MuLan \citep{huang2022mulan} are joint music-text representations that largely capture the semantic content of the underlying music and language description. On the other end, auto-encoder style models like WaveVAE \citep{WAVEVAE-peng2020nonautoregressiveneuraltexttospeech} or Music2Latent \citep{MUSIC2LATENT-pasini2024music2latentconsistencyautoencoderslatent} compress waveform into an n-dimensional continuous latent space that largely captures acoustic content. 
In both LM-based and diffusion-based approaches, the choice of representation for audio signals is crucial. Methods for encoding speech, environmental sounds, and music exist on a spectrum from high-level semantics to low-level acoustic representations. On the semantic end, models like CLAP \citep{CLAP-elizalde2022claplearningaudioconcepts} and MuLan \citep{huang2022mulan} use joint audio-text representations to capture the audio semantics of music and corresponding text descriptions. On the acoustic end, autoencoder-style models such as WaveVAE \citep{WAVEVAE-peng2020nonautoregressiveneuraltexttospeech} and Music2Latent \citep{MUSIC2LATENT-pasini2024music2latentconsistencyautoencoderslatent} compress waveforms into continuous latent spaces that capture the acoustic details.
Discrete neural audio codecs like SoundStream \citep{zeghidour2021soundstream}, Encodec \citep{ENCODEC-défossez2022highfidelityneuralaudio}, Descript Audio Codec \citep{DESCRIPT-kumar2023highfidelityaudiocompressionimproved}, HiFi-Codec \citep{HIFICODEC-yang2023hificodecgroupresidualvectorquantization}, and WavTokenizer \citep{WAVTOKENIZER-ji2024wavtokenizerefficientacousticdiscrete} capture acoustic information using a series of quantized codebooks with increasing levels of frequency details.
%Spectrogram-derived approaches such as AudioMAE \citep{AUDIO-MAE-huang2023maskedautoencoderslisten} and Wav2Vec-BERT \citep{WAV2VEC-BERT-chung2021w2vbertcombiningcontrastivelearning} are pretraining stages capturing varying mixtures of acoustic and semantic information suited to the downstream task. 
Spectrogram-derived approaches such as AudioMAE \citep{AUDIO-MAE-huang2023maskedautoencoderslisten} and Wav2Vec-BERT \citep{WAV2VEC-BERT-chung2021w2vbertcombiningcontrastivelearning} can be used as pre-training stages to capture a varying mixture of acoustic and semantic information suited to downstream tasks.
%Some approaches operate on spectrograms directly as part of a text-based in-painting framework like VoiceBox \citep{VOICEBOX-le2024voicebox} or AudioBox \citep{AUDIOBOX-vyas2023audioboxunifiedaudiogeneration} whilst others may even use aggressively down-sampled waveform in a diffusion-based framework \citep{NOISE2MUSIC-huang2023noise2musictextconditionedmusicgeneration} or VQ-VAE framework \citep{JUKEBOX-dhariwal2020jukeboxgenerativemodelmusic}.
Some methods, such as VoiceBox \citep{VOICEBOX-le2024voicebox} and AudioBox \citep{AUDIOBOX-vyas2023audioboxunifiedaudiogeneration}, directly manipulate spectrograms within a text-based inpainting framework, while others use auto-regressively down-sampled waveforms in diffusion-based \citep{NOISE2MUSIC-huang2023noise2musictextconditionedmusicgeneration} or VQ-VAE-based frameworks \citep{JUKEBOX-dhariwal2020jukeboxgenerativemodelmusic}.
%Depending on the task at hand, these approaches present varying levels of advantages and trade-offs. What type of representation is most suitable for music, methods to compress short term and long term features, mechanisms for encoding melodic and rhythmic domain knowledge and the trade-off between semantic and acoustic information continues to be active areas of work.

Each approach offers distinct advantages and limitations depending on the task. Ongoing research continues to explore optimal representations for music audio,  focusing on aspects such as compression methods that balance between short-term features and long-term coherence, disentanglement mechanisms for melodic, harmonic, and rhythmic characteristics, and the trade-offs between high-level semantic and low-level acoustic features.

%\paragraph{Evalutation metrics}
%\citep{EVAL1-yang2020evaluation} \citep{EVAL2-ji2020comprehensive}

\section{Method}
\label{sec:method}

\begin{figure}[ht]
    \centering
	\includegraphics[width=0.5\textwidth]{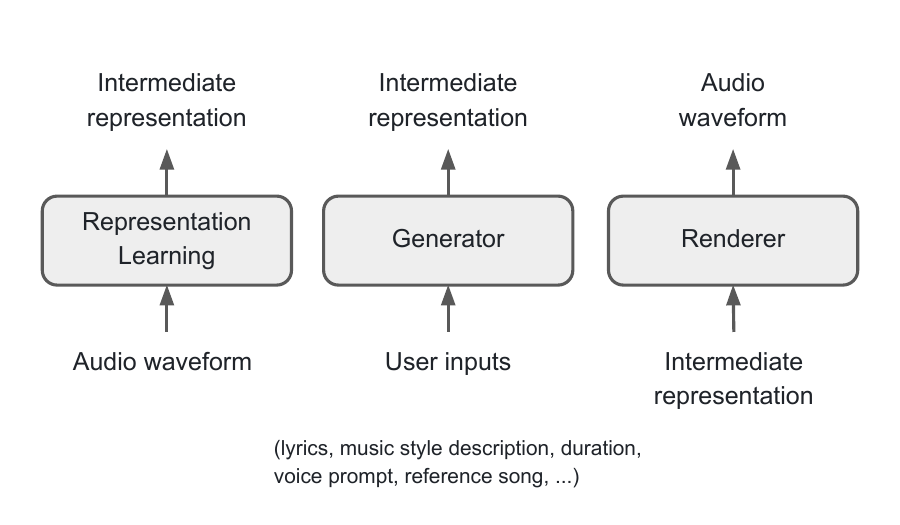}  
	\caption{An overview of Seed-Music framework.
 }
\label{fig:sys}
\end{figure}

Our music generation system consists of three core components as illustrated in \autoref{fig:sys}: a \textbf{Representation Learning module}, which compresses the raw audio waveform into the intermediate representation that serves as the foundation for training the subsequent components; a \textbf{Generator}, which processes various user control inputs and generates the corresponding intermediate representation; and a \textbf{Renderer}, which synthesizes high-quality audio waveform based on the intermediate representation from Generator. 

The primary design choice is the intermediate representation. As outlined in \autoref{sec:literature}, we identify three practical options: audio tokens, symbolic music tokens, and vocoder latents. The advantages and limitations of each are summarized in Table~\ref{tab:comparison_intermediate}.

\begin{table}[h]
\centering
\resizebox{\columnwidth}{!}{
\begin{tabular}{l|c|c|c|c}
\toprule
\textbf{Representation} & \textbf{Compression} & \textbf{Interpretability} & \textbf{Generator-Friendly} & \textbf{Renderer-Friendly} \\
\hline
Audio token     &  yes          & no  & yes         & maybe \\
\hline
Symbolic music  &  yes          & yes & yes         & no \\
\hline
Vocoder latent  &  maybe        & no  & maybe       & yes \\
\bottomrule
\end{tabular}
}
\caption{\label{tab:dit}\small Comparison of different intermediate representations.}
\label{tab:comparison_intermediate}
\end{table}

\begin{itemize}[leftmargin=2em]
\item \textbf{Audio tokens} are designed to efficiently encode both semantic and acoustic information at a much lower token rate than the audio sampling rate \citep{MUSICLM-agostinelli2023musiclmgeneratingmusictext,MUSICGEN-copet2024simplecontrollablemusicgeneration,JUKEBOX-dhariwal2020jukeboxgenerativemodelmusic}. When used with an auto-regressive LM based Generator, audio tokens serve as effective representations for connecting different modalities. However, their primary limitation lies in their lack of interpretability. Musical attributes such as vocal pronunciation, timbre, and pitch are embedded in a highly entangled format. 
Previous work \citep{AUDIOLM-borsos2023audiolmlanguagemodelingapproach} has explored how some audio tokens correspond to semantic features, while others capture acoustic aspects. This entanglement makes it challenging for the Generator to control specific elements of music, like melody and timbre, during audio token generation.

\item \textbf{Symbolic representations}, such as MIDI, ABC notation and MusicXML, are discrete and can be easily tokenized into a format compatible with LMs. 
Unlike audio tokens, symbolic representations are interpretable, allowing creators to read and modify them directly. However, their lack of acoustic details means the system has to rely heavily on the Renderer’s ability to generate nuanced acoustic characteristics for musical performance. Training such a Renderer requires large-scale datasets of paired audio and symbolic transcriptions, which are especially scarce for vocal music.

\item \textbf{Vocoder latents} from a variational auto-encoder serve as continuous intermediate representations, especially when used with diffusion models \citep{STABLEAUDIO-OPEN-evans2024stableaudioopen}. These latents capture more nuanced information compared to quantized audio tokens, allowing for a lighter Renderer in this pipeline. However, similar to audio tokens, vocoder latents are uninterpretable. Moreover, since vocoder latents are optimized for audio reconstruction, they may encode too much acoustic detail that is less useful for the Generator’s prediction tasks.
\end{itemize}

The selection of an intermediate representation depends on the specific downstream music generation and editing tasks. In the rest of this section, we present the technical details of our system design with these three intermediate representations and showcase their applications in Section \ref{sec:experiments}.

\subsection{Audio Token-based Pipeline}
\label{sec:audio-token}
\begin{figure}[h]
\centering
\includegraphics[width=1.0\textwidth]{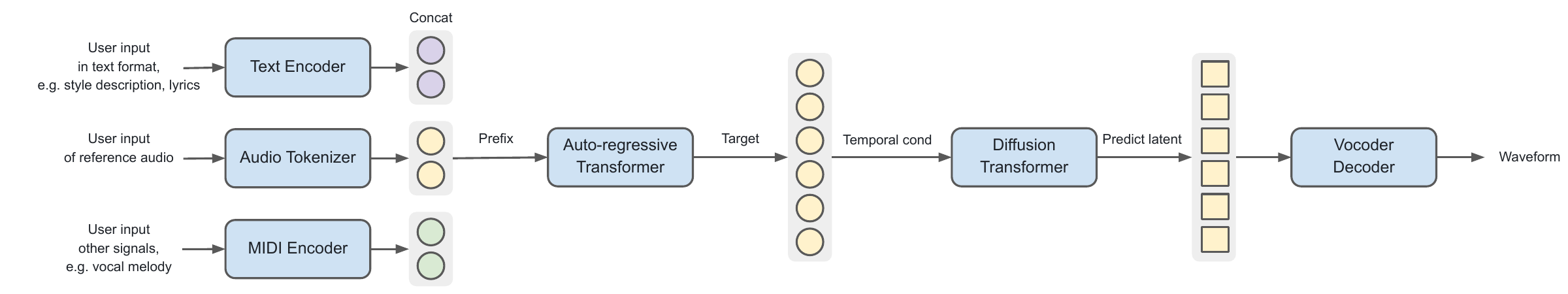}
\caption{\small Overview of the Seed-Music pipeline with audio token as intermediate representation. 
(1) Input embedders convert multi-modal controlling inputs, such as music style description, lyrics, reference audio, or music scores, into a prefix embedding sequence. (2) The auto-regressive LM generates a sequence of audio tokens. (3) The diffusion transformer model generates continuous vocoder latents. (4) The acoustic vocoder produces high-quality 44.1kHz stereo audio.}
\label{fig:token_sys}
\end{figure}

The audio token-based pipeline, as illustrated in Figure~\ref{fig:token_sys}, includes four building blocks: (1) an audio tokenizer, which converts raw music waveforms into low rate discrete tokens; (2) an auto-regressive LM (i.e. the Generator), which takes in user control inputs, convert them into prefix tokens, and predicts a sequence of target audio tokens; (3) a token diffusion model, which predicts the vocoder latents based on the audio tokens; and (4) an acoustic vocoder, which renders the final 44.1kHz stereo audio waveform. The token-to-latent diffusion module and latent-to-waveform vocoder module collectively form the token-to-waveform process, referred to as the Renderer. 

\paragraph{Audio tokenizer.}
The effectiveness of the audio tokenizer is critical to the success of this pipeline. The audio tokens embed key musical information from the original signals, such as melody, rhythm, harmony, phonemes, and instrument timbre. Our implementation is inspired by \cite{betker2023better}, \cite{wang2023neural}, and \cite{lajszczak2024base}, with further optimizations in architecture and training to achieve the following:
\begin{itemize}[leftmargin=2em]
\item High retention of essential information at a low compression rate, improving the training efficiency of the auto-regressive LM.
\item A balance between semantic and acoustic features, ensuring sufficient semantic details to optimize the Generator training while maintaining enough acoustic details for accurate waveform reconstruction by the Renderer. This trade-off between token generation and signal reconstruction \citep{TRADEOFF-blau2019rethinking} is carefully managed.
\end{itemize}

\paragraph{Generator.}
The auto-regressive LM generates audio tokens by conditioning on control signals that steer the generation towards the desired audio output. Each training example consists of paired annotations and audio, with the annotations converted into a sequence of embeddings that serves as the prefix for the LM. The handling of different control signal modalities is summarized as follows:
\begin{itemize}[leftmargin=2em]
    \item Categorical signals: Closed-vocabulary tags (e.g., music genre) are converted into categorical embeddings using a lookup table, while free-form text descriptions are processed using a general-purpose text encoder from MuLan~\citep{huang2022mulan}.
    \item Floating-point signals: Variables like melody note duration or song length are embedded using xVal encoding \citep{XVAL-golkar2023xval} to represent continuous numerical inputs.
    \item Lyrics signals: Lyrics are transformed into phoneme sequences to capture pronunciation, improving the model’s generalization to unseen words.
    \item Reference audio signals: The tokenizer extracts discrete token sequences from the reference audio, which are then mapped to continuous embeddings using a lookup table of the same size as the tokenizer’s codebook, or further aggregated into track-level embeddings. 
\end{itemize}    
During training, the model minimizes cross-entropy loss on a next-token prediction task using teacher forcing. At inference, user inputs are converted into prefix embeddings based on the specified modalities, and the audio tokens are generated auto-regressively.

\paragraph{Renderer.}
Once the auto-regressive LM generates the audio tokens, these tokens are processed by the Renderer to produce a rich, high-quality audio waveform. The Renderer is a cascaded system composed of two components: a Diffusion Transformer (DiT) and an acoustic vocoder, both of which are trained independently.
The DiT employs a standard architecture with stacked blocks of attention layers and multi-layer perceptrons (MLPs). Its objective is to reverse the diffusion process, predicting clean vocoder latents from noise by estimating the noise level at each step. The acoustic vocoder is the decoder from a low frame-rate VAE vocoder and follows designs similar to \citep{kumar2024high,lee2022bigvgan,cong2021glow,liu2021basis}. 
We found that structuring the vocoder latents as an information bottleneck within the cascaded system, combined with optimizing it with manageable model size and training time, results in superior audio quality and richer acoustic details compared to a single model that directly converts audio tokens into waveform.

\subsection{Symbolic Token-based Pipeline}
\label{sec:symbolic-token}

\begin{figure}[h]
\centering
\includegraphics[width=1\textwidth]{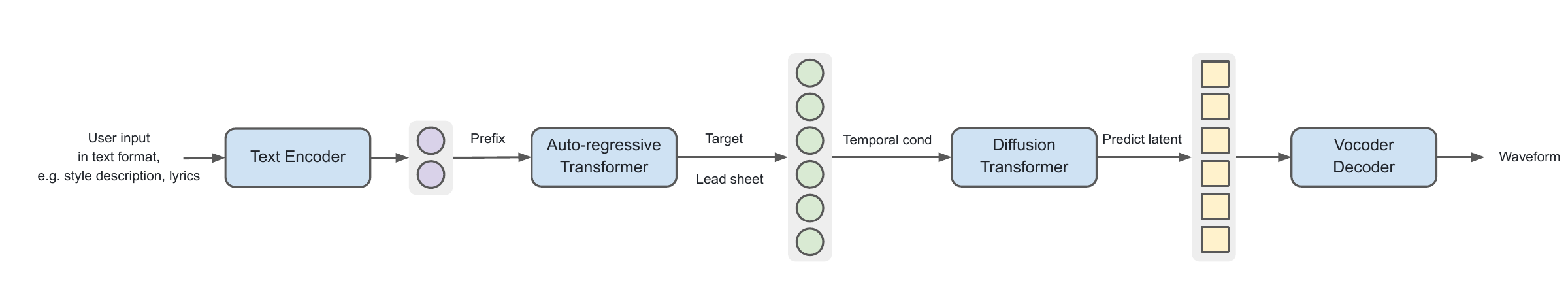}
\caption{\small Overview of the pipeline using symbolic tokens as the intermediate representation. (1) Conditioned on the user prompt, the auto-regressive LM generates the symbolic tokens corresponding to a lead sheet. (2) The diffusion transformer model generates continuous vocoder latents given the symbolic tokens. (3) The vocoder then generates the high-quality 44.1KHz stereo audio waveform.}
\label{fig:symbolic_sys}
\end{figure}

In contrast to the audio token-based pipeline, the symbolic token-based Generator, as shown in Figure~\ref{fig:symbolic_sys}, is designed to predict symbolic tokens for better interpretability, which is crucial for addressing musicians’ workflows in Seed-Music. 

Prior efforts have proposed algorithms for melody generation \citep{ju2021telemelody,zhang2023modeling}.
However, they lack explicit phoneme- and note-aligned information crucial for vocal music generation. 
Moreover, they remain limited to symbolic music generation without the capability for audio rendering. 
In a different line of research, there are task-specific prior works studying the approaches to steer music audio generation through musically interpretable conditions like harmony \citep{MUSICGEN-copet2024simplecontrollablemusicgeneration}, dynamics, and rhythm \citep{MUSIC-CONTROLNET-wu2023musiccontrolnetmultipletimevarying}. 
Inspired by how jazz musicians use lead sheets to outline a composition's melody, harmony and structure, we introduce ``lead sheet tokens'' as the symbolic music representation. We highlight the key components, benefits, and limitations of lead sheet tokens compared to audio tokens as follows.
\begin{itemize}[leftmargin=2em]
    \item To extract the symbolic features from audio for training the above system, we utilize in-house Music Information Retrieval (MIR) models, including beat tracking \citep{hung2022modeling}, key and chord detection \citep{lu2021spectnt}, structural section segmentation \citep{wang2022catch}, five-instrument MIDI transcription (i.e., vocals, piano, guitar, bass, and drums) \citep{TRANSCRIPTION-lu2023multitrack,TRANSCRIPTION-wang2024melrof}, and singing lyrics transcription. 
    The lead sheet tokens represent note-level details such as pitch, duration,  position within a bar, vocal phonemes aligned to notes, and track-level attributes like section, instrument, and tempo.
    \item The one-to-one mapping between lead sheet tokens and human-readable lead sheets allows creators to understand, edit, and interact with the musical scores directly. We experimented with different methods to generate lead sheet token sequences: REMI-style \citep{REMI-huang2020pop} and xVal \citep{XVAL-golkar2023xval}. 
    The REMI-style method interleaves instrument tracks into a quantized beat-based format, while xVal encodes onset and duration as continuous values. Although xVal-style encoding better follows our generative model's end product, the music performance, more closely, we found that the REMI-style one better suited for user interaction with musicians.
    \item Lead sheet tokens allow for the incorporation of human knowledge during both training and inference. For instance, music theory rules can be applied as constraints when predicting the next token in the sequence to enhance prediction accuracy.
    \item As the lead sheet tokens lack acoustic feature characterization, we need to scale up the token-to-latent diffusion model in the cascaded Renderer to achieve the same end-to-end performance as the audio token-based system.
\end{itemize}

\subsection{Vocoder Latent-based Pipeline}
\label{sec:vae-latent}

\begin{figure}[h]
\centering
\includegraphics[width=1\textwidth]{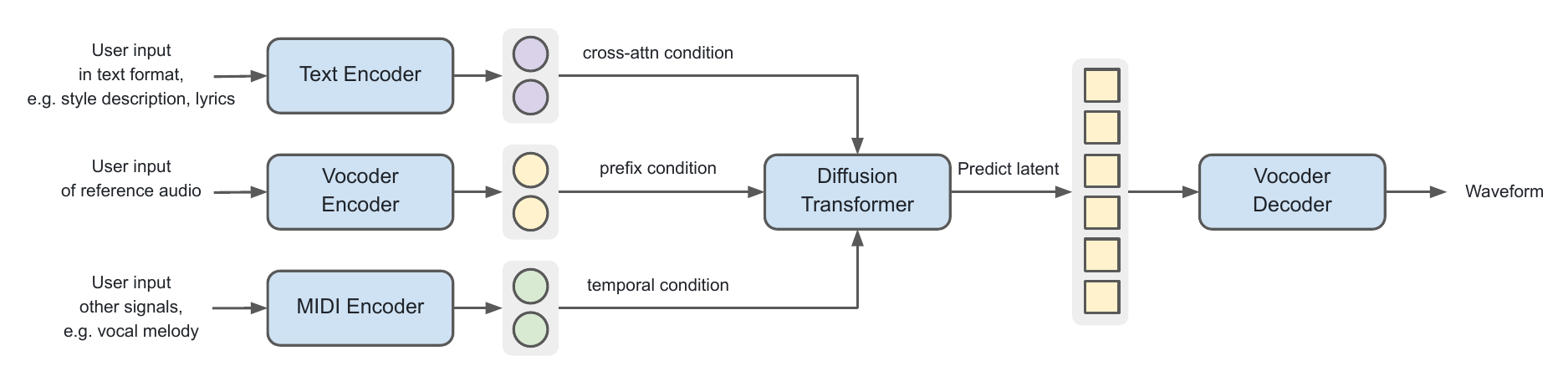}
\caption{\small Seed-Music pipeline with vocoder latents as intermediate representation. (1) Various input types are fed into DiT via cross-attention, prefix, or temporal conditioning. (2) The diffusion transformer model predicts the continuous vocoder latents. (3) The acoustic vocoder then produces high-quality 44.1kHz stereo audio.}
\label{fig:dit_sys}
\end{figure}

Previous works \citep{evans2024long,evans2024fast,levy2023controllable, LATENT-DIFFUSION-rombach2022highresolutionimagesynthesislatent} have shown that an efficient approach for the task of ``text-to-music'' is to directly predict the vocoder latents using a latent diffusion model. Similarly, we train a variational autoencoder (VAE) operating at a low latent frame rate, alongside a diffusion transformer (DiT) that maps conditional inputs to normalized, continuous vocoder latents, as illustrated in Figure~\ref{fig:dit_sys}.

Compared to the audio token-based pipeline (see Section~\ref{sec:audio-token}), the auto-regressive transformer module is omitted, although the architectures of the DiT and vocoder remain largely similar. To achieve comparable performance, the model size of each remaining module is scaled up.
In the auto-regressive approach, all conditioning inputs are encoded into tokens in the prefix sequence, which can result in an excessively long prefix that degrades performance when handling larger and more diverse inputs.
On the contrary, the vocoder latent-based design offers greater flexibility for incorporating a wider range of conditioning signals and supporting multi-channel inputs and outputs \citep{NOISE2MUSIC-huang2023noise2musictextconditionedmusicgeneration}. We summarized how different types of prompts are used as follows:
\begin{itemize}[leftmargin=2em]
    \item In-context conditioning in the vocoder latent space: This enables audio in-painting scenarios, such as audio continuation and editing. 
    \item In-context conditioning in the input noise space \citep{DIFFUSION-peebles2023scalablediffusionmodelstransformers}: For variable-length inputs like lyrics and style descriptions, cross-attention layers are applied at each transformer block to incorporate these inputs.
    \item Temporal inputs that span multiple tracks: Time-varying signals such as melody contours, intensity curves, and instrumental stems that are time-aligned conditioning inputs can be added at each step of the denoising process.
    \item Multi-channel outputs: Supported when multi-channel output examples are provided during training. For instance, the model can generate multiple musically distinct stems (e.g., vocals, bass, drums, and guitar), enabling downstream production scenarios like mashups and remixing. These stem-level training examples can be obtained from Music Source Separation (MSS).
\end{itemize}

\subsection{Model Training and Inference}
% Seed-Music is trained on large amounts of data to enable strong generalization and emergent capabilities. 
For all above mentioned pipelines, Seed-Music undergoes three training stages: pre-training, fine-tuning, and post-training similar to Seed-TTS and other text-based LMs. The pre-training stage aims to establish a better foundation for general music audio modeling. The fine-tuning stage consists of either data fine-tuning to enhance musicality, or instruction fine-tuning to improve controllability, interpretability, and interactivity for specific creation workflows. 

Post-training of Seed-Music is conducted through 
Reinforcement Learning (RL), which has proven to be an effective learning paradigm in text and image processing \citep{schulman2017proximal,rafailov2024direct,sutton1999policy,esser2024scaling,wallace2023diffusion}. Recent research has shown that Proximal Preference Optimization (PPO) can be extended to music and speech generation \citep{cideron2024musicrl, zhang2024speechalign}. 

Inspired by these findings, we explore RL methods \citep{ahmadian2024back,prabhavalkar2018minimum,wang2024transforming,sutton1999policy,schulman2017proximal} 
%to enhance the output's adherence to different input control signals and improve musicality.
to improve the alignment of generated output with various input control signals and enhance musicality.
Reward models we considered include: 
%the edit distance between the original lyrics prompt and the lyrics transcription obtained from the generated audio; genre prediction accuracy comparing the genre input against the detected genre of the audio output; the match between a song structure prompt and the song structure detected from generated audio.
the edit distance between the original lyrics prompt and the lyrics transcription extracted from the generated audio,
the genre prediction accuracy by comparing the input genre with the detected genre of the audio output, 
and the match between a song structure prompt and the detected structure in the generated audio.
Additional reward models based on tempo, instrumentation, audio references, and user voice prompts can be used to dictate what musical attributes are emphasized in the generation output. 
Moreover, incorporating human feedback \citep{RLHF-ouyang2022training} can produce reward models that capture subtle user preferences beyond the above objective metrics. We leave the thorough study of RL for future work.

During inference, the choice of sample decoding scheme plays a critical role in both output quality and stability for auto-regressive and diffusion models. We observed that carefully tuning classifier-free guidance \citep{CFG-ho2022classifier, CFG-sanchez2023stay} is essential to ensure musicality and adherence to prompts. To reduce latency, we apply model distillation \citep{CONSISTENCY-song2023consistency} to minimize the number of iteration steps required by the DiT models. Additionally, we developed a streaming decoding scheme, allowing audio to be streamed while the auto-regressive model continues generating the token sequence. 
%
% For diffusion transformer, noise schedule also plays an important \citet{NOISE-lin2024common}. 

% Finally, to improve the latency of the cascaded generation system, we applied model distillation following \cite{CONSISTENCY-song2023consistency} to reduce the number of iterations required for the diffusion model inference; and taking advantage of the auto-regressive decoding of LM, we also implemented streaming decoding scheme to run the language model and diffusion model inference in parallel, so that the music audio is streamed while the language model continues to generate the audio token sequence.

\section{Experiments}
\label{sec:experiments}

In this section, we showcase four applications powered by our model's capabilities: Lyrics2Song (Section~\ref{ssec: lyrics2song}), Lyrics2Leadsheet2Song (Section~\ref{ssec: lyrics2leadsheet2song}), MusicEDiT (Section~\ref{ssec: music-editing}), and Zero-shot Singing Voice Conversion (Section~\ref{ssec: voice-conversion}).

In \textbf{Lyrics2Song}, we introduce a vocal music generation system that produces performance-quality music with vocals based on user-provided lyrics and music style inputs.
\textbf{Lyrics2Leadsheet2Song} builds on the Lyrics2Song system by incorporating symbolic music representation for enhanced interpretability. This process additionally generates a lead sheet, where users can access and adjust the melody and rhythm, allowing for finer control over the final audio output.
\textbf{MusicEDiT} explores a diffusion-based in-painting system that enables users to edit the lyrics and melodies of an existing music audio piece. This serves as a post-production tool for modifying the vocals of a song. 
In \textbf{Zero-shot Singing Voice Conversion}, we offer a solution that allows users to modify the timbre of vocals in an existing audio based on their own voice with minimal reference data. This application facilitates vocal personalization with a low preparation cost.
For each of these aforementioned applications, we discuss the design choices related to intermediate representations, model architecture, and other configurations that optimize the system for its respective use case.

\subsection{Lyrics2Song}
\label{ssec: lyrics2song}
Lyrics2Song generates vocal music performance conditioned on user-provided music style descriptions and lyrics with section tags (e.g., `verse', `chorus', and `bridge') \citep{wang2022musfa}. This task utilizes the audio token-based pipeline (see Section~\ref{sec:audio-token}), leveraging tokenization and auto-regressive techniques to align multi-modal data (i.e., lyrics, styles, tags, and audio) and enable streaming decoding for fast, responsive interactions.
%generates vocal music performances conditioned on lyrics and music style description. The audio token-based pipeline (see Section~\ref{sec:audio-token}) is particularly developed for this task. The use of tokenization and auto-regressive approaches allows for multi-modal alignment (i.e., lyrics, tags, and audio) and streaming decoding scenario which further facilitates fast and responsive interactions.

%Our system is capable of cohesive generation for both short-form audio clips \footnote{\url{https://team.doubao.com/seed-music/shortform-audio-generation}} and long-form full length tracks \footnote{\url{https://team.doubao.com/seed-music/longform-audio-generation}}.
%The audio samples demonstrate many desirable musical characteristics, including expressive and dynamic vocal performances, interesting and pleasing melodies as well as instrumentals that reflect a diverse range of genres and moods. 

This system supports both cohesive short-form audio clip generation\footnote{\url{https://team.doubao.com/seed-music/shortform-audio-generation}} and full-length track production\footnote{\url{https://team.doubao.com/seed-music/longform-audio-generation}}. The generated audio showcases expressive and dynamic vocal performances with engaging melodies as well as instrumentals that span a wide variety of instrumentation and genres, delivering a mature level of musicality.

\paragraph{Vocal music generation with audio reference.} 
In addition to style descriptions, our system also supports audio input as a prompt to guide music generation. The listening examples\footnote{\url{https://team.doubao.com/seed-music/audio-prompting}} demonstrate how outputs are generated by referencing the musical styles of the audio prompts. Since describing desired music with text or tags can be less intuitive for novice users, audio prompts provide a more effective way to communicate musical intent. 

%Our system supports two modes of audio prompting. In ``continuation mode'', the sequence of audio tokens extracted from the audio prompt is concatenated in the prefix to continue the auto-regressive generation. This ensures the generated output preservation of structural, melodic, and sonic similarities to the reference audio prompt. 
%In ``remix mode'', the input audio is extracted into an embedding vector in a pre-trained joint embedding space of text and audio \citep{huang2022mulan}, so that it can be concatenated in the prefix for the auto-regressive LM. The embedding vector summarizes the {\it global} characteristics of the audio prompt.

Our system supports two modes of audio prompting: \emph{continuation mode} and \emph{remix mode}. In continuation mode, audio tokens extracted from the audio reference are concatenated in the prefix to continue auto-regressive generation, ensuring strong structural, melodic, and sonic similarities to the reference. In remix mode, the audio reference is converted into an embedding vector within a pre-trained joint text-audio embedding space \citep{huang2022mulan}. This embedding, which summarizes the global characteristics of the audio reference, is then incorporated into the prefix to guide the generation, allowing the generated audio to adopt different styles.

In both modes, our model demonstrates a strong ability to maintain coherence between the input lyrics and the inherent lyrics in the audio reference, even without the aid of automatic lyrics transcription. When the input lyrics are structurally and semantically similar to those in the audio reference, the model tends to resemble the melody and structure from the reference. However, when the input lyrics differ significantly in style (e.g., language, structure, rhyme), the coherence weakens. Despite this, the model effectively maintains natural rhythm patterns, instrumentation, vocal quality, and overall musical motifs.

\textbf{Instrumental music generation.} 
Although the audio token-based pipeline is primarily designed for vocal music generation, it also supports instrumental music generation if the lyrics input contains only section tags without text. We provide several generated instrumental examples\footnote{\url{https://team.doubao.com/seed-music/instrumental-music-generation}} in a wide variety of styles, with each section unfolding at the specified time and demonstrating clear structural transitions between sections (e.g., verse to chorus).

\textbf{Evaluation metrics.} 
We used the following quantitative metrics to assess generation quality during development. These metrics were also repurposed as reward models in the Reinforcement Learning process for the auto-regressive LM.
\begin{itemize}[leftmargin=2em]
    \item \textbf{Word error rate (WER)}: We use in-house singing lyrics transcription model, which supports both English and Mandarin Chinese languages, to transcribe the generated audio and compute the word (or Pinyin) error rate relative to the lyrics prompt. While useful, WER is not a perfect measure of vocal quality in music due to factors like elongated vowels, consonants, pitch variations, and non-speech-like rhythms in sung words. These characteristics can introduce noise into the WER calculation.
    \item \textbf{Music tagging performance}: To evaluate the alignment between the generated audio and input style descriptions, we use in-house music tagging and structural segmentation models to predict high-level musical attributes from the generated audio, including genre, mood, vocal timbre, vocal gender, and structural sections. These predicted attributes are then compared to the input style descriptions and section tags, with average precision scores serving as the quantitative metric for relevance.
\end{itemize}

For qualitative evaluation, we use the Comparative Mean Opinion Score (CMOS), based on feedback from a team of musically-trained raters. We define the following three dimensions for assessment:
\begin{itemize}[leftmargin=2em]
    \item \textbf{Musicality} evaluates musical attributes, including novelty of vocal melodies, appropriate use of harmony, idiomatic musical forms (e.g. theme, variation), coherent structure, suitable chord progressions, characteristic rhythmic patterns, and well-rounded instrumentation.
    % \item \textbf{Musicality} measures musically semantic attributes and was decomposed into interesting vocal melodies, appropriate use of harmony, presence of idiomatic musical forms like theme and variation, plausible structure, suitable chord progressions, characteristic rhythmic patterns, and full-bodied instrumentation arrangement. 
    \item \textbf{Audio quality} assesses acoustic characteristics such as vocal clarity, instrument realism, detail across the frequency spectrum, and the sharpness of drum transients and onsets. Raters also consider any unwanted audio artifacts, such as distortion, muffling, or missing energy in certain frequency bands.
    %\item \textbf{Audio Quality} measures acoustic attributes including the clarity of vocals, realism of instruments, detail across the frequency spectrum, and transients of drums and onsets. Raters also consider unwanted audio artifacts like distortion, muffled audio, and missing energies in certain frequency bands.
    \item \textbf{Prompt adherence} measures how closely the generated audio aligns with the input lyrics and style prompts.
    %\item \textbf{Prompt Adherence} measures how accurate the generated audio resembles the input lyrics and style prompts. 
\end{itemize}

In the speech domain, benchmark datasets are established to evaluate TTS systems using metrics like WER and Automatic Speaker Verification (ASV). However, there are no equivalent benchmarks for music generation that provide quantitative scores. Additionally, it is important to note that \emph{musicality}—a key factor in assessing music generation quality—is highly subjective and challenging to quantify with objective metrics. We encourage readers to listen to the provided audio demos to better assess the quality of our system.

%We note that musicality, one of the most important aspect in justify the quality of music generation, is highly subjective and nearly impossible to be quantified by an objective metric. We encourage readers to directly listen to the provided audio demos to judge the quality of our system, which is more meaningful than the quantitative numbers reported.

%We hope this article can inspire subsequent research in vocal music generation.
%We note the challenge in presenting quantitative and qualitative statistics that can be compared against for music generation, and thus 
% and may choose to report metrics with matching models for meaningful benchmarking in the future. 

\textbf{Audio tokens versus vocoder latents.} 
%We have also conducted Lyrics2Song experiments with a diffusion-based pipeline as outlined in Section~\ref{sec:vae-latent}, and achieved on-par end-to-end performance. 
%Nevertheless, we find the auto-regressive LM-based approach to be naturally suited for interactive use cases. Its causal nature enables streaming solutions like those described in Section~\ref{sec:audio-token} to provide a near real-time experience and allows future integration with multi-modal models \citep{MU-LLAMA-liu2023musicunderstandingllamaadvancing}. 
We also conducted Lyrics2Song experiments using the vocoder latent-based pipeline (see Section~\ref{sec:vae-latent}), achieving performance comparable to the audio token-based pipeline (see Section~\ref{sec:audio-token}). However, we find the auto-regressive LM to be inherently better suited for interactive applications than the diffusion model. Its causal architecture enables streaming solutions that provide a near real-time experience, while also allowing for future integration with multi-modal models \citep{MU-LLAMA-liu2023musicunderstandingllamaadvancing}.

\subsection{Lyrics2Leadsheet2Song}
\label{ssec: lyrics2leadsheet2song}

The Lyrics2Leadsheet2Song pipeline is a two-step process for achieving the Lyrics2Song task: \textbf{Lyrics2Leadsheet} and \textbf{Leadsheet2Song}. In the first step, lead sheet tokens are generated from the input lyrics and style descriptions. In the second step, music audio is produced from the lead sheet tokens. The overall pipeline is illustrated in Section~\ref{sec:symbolic-token}. Lead sheet tokens enable user involvement in the generation process by allowing edits to melody, chords, instrumentation, and tempo before the final rendering.

\begin{figure}[ht]
    %\centering	
    \includegraphics[width=1\textwidth]{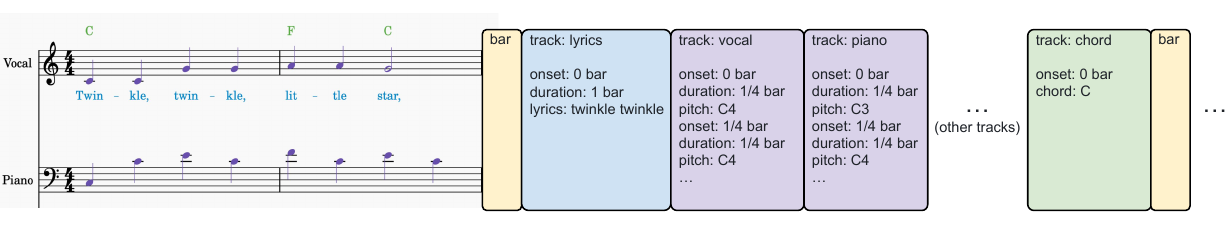}  
	\caption{Illustration of the REMI-style symbolic music encoding scheme.}
\label{fig:sympac_encodec}
\end{figure}

\textbf{Lyrics2Leadsheet.} 
We developed a rule-based symbolic music encoding scheme based on \citep{SYMPAC-chen2024sympac} to encode the symbolic features of a music audio piece into a sequence of lead sheet tokens. As illustrated in \autoref{fig:sympac_encodec}, the scheme encodes lyrics and various musical events. It recognizes eight types of events: lyric phoneme, bar, chord, vocal note, bass note, piano note, guitar note, and drum note. Each type of events, except for `bar', is represented as a distinct ``track'' in the lead sheet tokens. Bar events define the basic temporal structure, with tracks interleaved on a bar-by-bar basis. For each event (e.g., phoneme, note, chord) within a track, we encode onset, duration, and pitch values when applicable.

As described in Section~\ref{sec:symbolic-token}, the Generator is trained using lead sheet tokens extracted by our in-house MIR models. Here are some examples\footnote{\url{https://team.doubao.com/seed-music/lyrics-to-leadsheet}} demonstrating how the auto-regressive LM predicts phoneme-aligned notes with genre-appropriate melodies and rhythms according to the input lyrics and style prompts.

\textbf{Leadsheet2Song.} 
The pipeline of Leadsheet2Song involves rendering a full audio mix from a sequence of lead sheet tokens. In the demo examples\footnote{\url{https://team.doubao.com/seed-music/leadsheet-to-song}}, we showcase how the generated vocal music follows the vocal melodies, phonemes, rhythms, chord progressions, and instrumental notes from the given lead sheet tokens. The Renderer effectively generates the nuances of natural and expressive music performance across multiple instruments, providing professionals with a powerful tool to quickly review audio results without the need for meticulous parameter adjustments in synthesizers.

\textbf{Leadsheet2Vocals for singing voice synthesis.} 
Rather than producing a full audio mix, the Lyrics2Leadsheet2Song system can be configured to generate individual stems, including vocals, drums, bass, piano, and guitar, in both symbolic notation as well as rendered audio. 
Singing Voice Synthesis (SVS) is one application of this system, where the model is set to output only vocal stem, as demonstrated in these examples\footnote{\url{https://team.doubao.com/seed-music/leadsheet-to-vocals}}.

\subsection{Music Editing}
\label{ssec: music-editing}
%In this section, we consider the task of editing a given music audio as a post-production process. The non-auto-regressive nature of diffusion-based approach described in Section~\ref{sec:vae-latent} makes it better suited for editing tasks. 
In this section, we explore music audio editing as a post-production process. The non-casual nature of the diffusion-based approach, described in Section~\ref{sec:vae-latent}, makes it particularly well-suited for such tasks.
%For example, in text-conditioned in-painting, a diffusion module’s visibility of context before and after the masked audio segments guarantees smoother transitions \citep{AUDIT-wang2023auditaudioeditingfollowing}.
%We formulate the problem as a lead sheet conditioned in-painting task to train the Diffusion Tranformer model. Upon inference, the modified lead sheet is used as conditioning input, and the segment of audio corresponding to the modified part of the lead sheet is masked and re-generated.
For example, in text-conditioned in-painting, a diffusion model’s ability to access context before and after masked audio segments ensures smoother transitions \citep{AUDIT-wang2023auditaudioeditingfollowing}. We frame this as a lead sheet-conditioned in-painting task to train the DiT model. During inference, the modified lead sheet serves as conditioning input, and the audio segment corresponding to the altered part of the lead sheet is masked and re-generated.

%In these listening examples\footnote{\url{https://team.doubao.com/seed-music/editing-lyrics}}, we show the system's ability to precisely change the sung lyrics while keeping the melody and backing tracks unchanged. We demonstrate this using English and Mandarin Chinese singing voices. In addition to changing the lyric words in the same language, there are cases where lyrics are sung in alternating languages. Other than that, in these demos\footnote{\url{https://team.doubao.com/seed-music/editing-melody}}, we show that users can also precisely modify the melody within a given time segment while keeping the lyrics, remaining melody, and backing track unchanged. We are particularly excited by this new paradigm of ``generative audio editing'', which retains the musical performance and essential characteristics of the original track. Such post-production editing scenarios would not be possible in the past without re-recording the vocals alone with the original instrumental stems.
%These are post-production edits that historically would have required original project stems and MIDI tracks.
In these listening examples\footnote{\url{https://team.doubao.com/seed-music/editing-lyrics}}, we demonstrate the system’s ability to precisely modify sung lyrics while preserving the melody and backing tracks using both English and Mandarin Chinese singing voices. 
In some cases, the lyrics are altered within the same language, while in others, the system allows for alternating between languages. Moreover, in these examples\footnote{\url{https://team.doubao.com/seed-music/editing-melody}}, we demonstrate how users can precisely adjust the melody within a specified time segment, while keeping the lyrics, the rest of the melody, and the backing track unchanged. This new ``generative audio editing'' paradigm excites us, as it retains the musical performance and essential qualities of the original track—something that was previously complex or nearly impossible without re-recording the vocals along with the original instrumental stems.

% \textbf{Evaluation Metrics} We use identical quantitative and qualitative metrics as in section 4.1.

\subsection{Zero-shot Singing Voice Conversion}
\label{ssec: voice-conversion}
One of the most intuitive ways for creators to edit vocal music is by converting the vocal timbre to match their own voice. This section explores the singing Voice Conversion (VC) system as the final component of the Seed-Music suite. Although our singing VC method shares similarities with the speech VC introduced in Seed-TTS \citep{SEED-TTS-anastassiou2024seedttsfamilyhighqualityversatile}, voice cloning and conversion in the context of vocal music generation present greater challenges compared to the speech domain \citep{arik2018neural}:
%In the context of Seed-Music, singing voice conversion (SVC) is the final piece of the suite that bridges a creator's own voice with the generative systems. 
\begin{itemize}[leftmargin=2em]
    \item \textbf{Vocal mixture}: 
    Vocal music typically consists of vocals and background instrumental tracks, with strong coherence between them in terms of harmony and rhythm. In contrast, speech signals often contain background environmental sounds unrelated to the speech content. While modern MSS models can isolate vocals, they often introduce artifacts that degrade quality. 
    %when conventional VC techniques are applied to these isolated vocals. 
    Our goal is to develop a scalable system capable of directly processing the mixture of vocals and background tracks without relying on MSS, thereby avoiding these artifacts.
    \item \textbf{\textbf{Vocal range}}: The pitch range of singing voices is much wider than that of speech. For zero-shot singing VC, the system must generalize the pitch range of the reference voices to the synthesized singing voices, requiring strong robustness from the model.
    \item \textbf{\textbf{Vocal technique}}: Singing voices are highly expressive and involve far more techniques than speech. The same vocalist can sound vastly different when performing in styles like operatic mezzo-soprano, musical theater belting, or jazz scatting. A singing VC system must accurately capture and reproduce these expressive techniques, in addition to handling regular speech attributes like clear pronunciation and prosody.
    \item \textbf{\textbf{Singing versus speech reference}}: In VC applications, users typically provide speech as their reference voice, whether for speech or singing voice synthesis. Our system is specifically designed to accept reference voices regardless of whether they are speech or singing. It can effectively perform singing VC using a short speech clip as the reference.
    \item \textbf{Amateur versus professional singing}: 
    There is significantly less paired data available for amateur versus professional singing compared to speech VC data. This makes singing VC particularly challenging, as the model has to adapt to non-professional singing inputs and convert them into professional-quality performances. For instance, if users provide singing references that are out of tune, the singing VC system must not only capture the tone of their voices but also correct the pitch.
    %Speech benefits from a relatively easy procedure for finding paired data between an input speaker and target speaker e.g. recording different voices reading the same passage. However, no such large scale data exists for modeling an amateur singer and professional singer. A vocal music VC system must translate non-professional singing into an improved performance, e.g. if the end-user provides out-of-tune singing, the VC result should retain the tone of voice but actually be in tune.
\end{itemize}
The listening examples\footnote{\url{https://team.doubao.com/seed-music/singing-voice-conversion}} demonstrate how our singing VC system performs across different scenarios.
The quality of the outcomes depends largely on the similarity between the reference voices and the target singing signals. For instance, when both are male voices singing in English, the results are optimal. However, handling cross-gender and cross-language cases is more challenging, often leading to issues such as artifacts, distortions, and inconsistencies in pronunciation.

%Naturally, when the input reference vocal and target conversion vocal are similar, i.e. both are male voices singing in English, the results are the best. Other combinations are more challenging, such as when the input reference vocal is a female \textit{speaking} (not singing) in Mandarin and the target conversion vocal is a female singing in Mandarin or even harder still, singing in English. The listening examples\footnote{\url{https://team.doubao.com/seed-music/singing-voice-conversion}} demonstrate how our VC system is able to handle these combinations.

\section{Conclusion}
\label{sec:conclusion}
In this report, we have introduced Seed-Music, a comprehensive suite of music generation and editing systems designed to support diverse music creation workflows. We have also demonstrated how the system generates high-quality vocal music conditioned on multi-modal inputs, including lyrics, style descriptions, audio references, music scores, and voice prompts. Our unified framework addresses various use cases using three intermediate representations (i.e., audio tokens, lead sheet tokens, and vocoder latents) and their associated pipelines, providing users with flexible tools to move from ideation to generation and editing. 
%The versatility of this unified framework opens up exciting possibilities for future developments, such as multi-stem generation and enhanced multi-modal conditioning signals.

%In this report, we have introduced Seed-Music, a comprehensive suite of music generation and editing systems for music creation workflows. We have demonstrated the generation of high-quality vocal music conditioned on multi-modal inputs such as lyrics, style descriptions, audio references, music score, and voice prompts. 
%Our approach tackles different music creation workflows with three intermediate representations and their associated pipelines, offering various functionalities for users to engage from ideation to generation and editing. The great flexibility of such a unified framework sheds light on many exciting future works, including multi-stem generation and multi-modal conditioning signals. 
%
%We proposed an interpretable lead sheet tokenizer, which empowered precise audio editing workflows.
%

% \textbf{Applications}
%From application point of view, Seed-Music lowers the barriers to artistic creation and musical expression. We strongly believe the demos shown in this report can empower creators across the spectrum of novices and professionals. For example, the coupling of text-to-music pipelines with zero-shot singing voice conversion is intended to engage novices more deeply into the creative process. Rather than manipulating music from a distance, novices bring their own unique voices and identities to participate in creative ideation. 
From application perspectives, Seed-Music lowers the barriers to artistic creation and musical expression. We believe the demos shown in this report can empower a wide range of creators, from novices to professionals. For example, the integration of text-to-music system with zero-shot singing voice conversion allows novices to engage more deeply in the creative process. Instead of interacting with music from a distance, novices can bring their own unique voices and identities into the process, enhancing creative ideation.

%Music is also integral to many complementary mediums including short-form video, long-form cinema, games, and AR/VR experiences. Real-time conditioning and rendering of generative music opens up entirely new interactions beyond traditional playback of audio files. We imagine entirely new artistic mediums where generative music responds to conditioning signals beyond text but in-game storylines and visual art-style. 
Music is also a key component of complementary media such as short-form videos, films, games, and AR/VR experiences. For more applications, real-time conditioning and rendering of generative music introduce entirely new forms of interaction, going beyond traditional audio playback. We envision new artistic mediums where generative music responds to conditioning signals not only from text but also from in-game narratives and visual art styles.

%For professionals, the proposed lead sheet tokens were engineered from the ground-up to respect and integrate into workflows of musicians, composers, vocalists and artists. We believe the lead sheet tokens can evolve to become the symbolic standard for music language models, similar to today’s industry-standard MIDI for traditional music making. Musicians and producers would be able to leverage the power of generative models through familiar means of control over melodic and rhythmic elements. Moreover, the ability to edit and manipulate recorded music in ways that preserve music semantics is a meaningful time-saving and cost-saving capability across the industry. We are also excited by future stem-based generation and editing workflows beyond the vocal track. Professionals can audition musical ideas with greater speed and experience higher chances of landing on ``happy accidents'' crucial to the creative process. 
For professionals, the proposed lead sheet tokens are designed to integrate seamlessly into the workflows of musicians, composers, vocalists, and artists. We believe these tokens have the potential to evolve into a symbolic standard for music language models, much like MIDI has for traditional music production. Musicians and producers could harness the power of generative models while maintaining familiar control over melodic, harmonic, and rhythmic elements. Furthermore, the ability to edit and manipulate recorded music while preserving its musical semantics offers significant time and cost savings for the industry. We are particularly excited about future developments in stem-based generation and editing, which would extend beyond vocal tracks. These capabilities will allow professionals to explore musical ideas more efficiently, increasing the likelihood of discovering the ``happy accidents'' that are often pivotal to the creative process.

% \textbf{Limitations} Despite its exciting capabilities, we note some of Seed-Music's current limitations:
% \begin{itemize}[leftmargin=2em]
%     \item The generated vocal track tends to lack the polish of modern music production such as reverb and delay. 
%     \item Instrumentals can sometimes be quiet relative to the vocal. When generating solo instrumentals, we note how transient and percussive details are much clearer than ones combining vocals and backing instrumentals. 
%     \item Perceptive listeners will hear how diffusion-based editing pipelines introduce small noisy artifacts in the areas where lyrics and melodies were edited. 
%     \item Singing voice synthesis begins to suffer when the vocal technique is intended to be a cross between spoken word and melodic singing. 
%     \item There is still much room for improvement for zero-shot voice conversion on plain speech compared to singing references. 
% \end{itemize}

\newpage
\section{Ethics and Safety}
\label{sec:ethics-safety}
We firmly believe that AI technologies should support, not disrupt, the livelihoods of musicians and artists. AI should serve as a tool for artistic expression, as true art always stems from human intention. Our goal is to present this technology as an opportunity to advance the music industry by lowering barriers to entry, offering smarter, faster editing tools, generating new and exciting sounds, and opening up new possibilities for artistic exploration.

% The training and dissemination of generative models should respect the copyright and ownership of Professionally Generated Content (PGC) and User Generated Content (UGC). 

\textbf{Ethics}
We recognize that AI tools are inherently prone to bias, and our goal is to provide a tool that stays neutral and benefits everyone. To achieve this, we aim to offer a wide range of control elements that help minimize preexisting biases. By returning artistic choices to users, we believe we can promote equality, preserve creativity, and enhance the value of their work. With these priorities in mind, we hope our breakthroughs in lead sheet tokens highlight our commitment to empowering musicians and fostering human creativity through AI.

\textbf{Safety}
In the case of vocal music, we recognize how the singing voice evokes one of the strongest expressions of individual identity. To safeguard against the misuse of this technology in impersonating others, we adopt a process similar to the safety measures laid out in Seed-TTS. This involves a multi-step verification method for spoken content and voice to ensure the enrollment of audio tokens contains only the voice of authorized users. We also implement a multi-level water-marking scheme and duplication checks across the generative process.

Modern systems for music generation may fundamentally reshape culture and the relationship between artistic creation and consumption. We are confident that, with strong consensus between stakeholders, these technologies will and revolutionize music creation workflow and benefit music novices, professionals, and listeners alike. 

\section{Acknowledgement}
\paragraph{Authors} (Alphabetical Order)
\begin{adjustwidth}{-0.2cm}{-1.2cm}
\begin{multicols}{5}
\noindent
Ye Bai \\
Haonan Chen \\
Jitong Chen  \\
Zhuo Chen  \\
Yi Deng  \\
Xiaohong Dong  \\
Lamtharn Hantrakul\\
Weituo Hao  \\
Qingqing Huang \\
Zhongyi Huang  \\
Dongya Jia  \\
Feihu La  \\
Duc Le  \\
Bochen Li  \\
Chumin Li  \\
Hui Li  \\
Xingxing Li  \\
Shouda Liu  \\
Wei-Tsung Lu  \\
Yiqing Lu  \\
Andrew Shaw  \\
Janne Spijkervet  \\
Yakun Sun  \\
Bo Wang  \\
Ju-Chiang Wang  \\
Yuping Wang  \\
Yuxuan Wang  \\
Ling Xu  \\
Yifeng Yang \\
Chao Yao \\
Shuo Zhang  \\
Yang Zhang  \\
Yilin Zhang  \\
Hang Zhao  \\
Ziyi Zhao  \\
Dejian Zhong  \\
Shicen Zhou  \\
Pei Zou
\end{multicols}
\end{adjustwidth}

We extend our gratitude to the larger Seed team whose dedication and expertise were vital to the success of this project. Special thanks to our engineering team for their technical prowess; our data teams, whose diligent efforts in data collection, annotation, and processing were indispensable; our project operation team for providing logistical guidance; our evaluation team for their rigorous testing and insightful feedback; and also the Seed-TTS and Seed-ASR teams for valuable knowledge sharing. Their contributions have been instrumental to Seed-Music (no pun intended).

\newpage
\bibliographystyle{unsrtnat}
\bibliography{sample}

\end{document}